\documentclass[useAMS]{mn2e}

\input{epsf}

\title{Linking radio to infrared: a radio source count model}
\author[A. J. King and M. Rowan-Robinson]
       {A. J. King\thanks{E-mail: alex.king@imperial.ac.uk} and M. Rowan-Robinson\\
        Astrophysics Group, Blackett Laboratory, Imperial College London, Prince Consort Road, London SW7 2AZ, UK}
\date{Accepted ????
      Received ????;
      in original form 2003 October 8}

\pagerange{\pageref{firstpage}--\pageref{lastpage}}
\pubyear{2003}

\begin{document}
\maketitle

\label{firstpage}

\begin{abstract}

We investigate the populations and evolution of normal and active galaxies by extending the infrared source count model of Rowan-Robinson (2001) into the radio. The FIR-radio correlation is used to extend the SEDs to the radio region and it is assumed that there are two distinct populations of quasar: radio loud and radio quiet. It is found that the radio luminosity function and source count data are best fit when the fraction of quasars that are radio loud is allowed to increase with optical luminosity. This implications for this are discussed, as are the possible causes for the variation in radio loud fraction.

\end{abstract}



\section{Introduction}
\label{sec:intro}

Source counts in the radio region of the spectrum divide into two general populations: a faint population of star-forming galaxies and a bright population of AGN-powered sources. This division was first revealed by the upturn in slope of the 1.4 GHz source counts, and has since been confirmed by spectroscopic and multicolour studies (e.g. Windhorst et al. 1985; Benn et al. 1993; Richards et al. 1998). The emission from bright radio sources ($S_{1.4 GHz} > 1$ mJy) is believed to be dominated by synchrotron radiation from jets powered by the accretion of matter onto an AGN, whereas the emission from faint sources ($S_{1.4 GHz} < 1$ mJy) is dominated by processes associated with young, massive stars (Thuan \& Condon 1987; Benn et al. 1993), hence is connected to the star formation rate (though some of these sources may also contain weak AGNs).

It is possible to investigate the nature of radio sources using source count modelling. This technique attempts to predict the observed number-flux relation using a local luminosity function and certain assumptions about the galaxy types, their relative contributions to the counts and their evolution. These models, in the radio, are typically based on the 1.4 GHz luminosity function (e.g. Condon 1984; Hopkins et al. 1998). These are able to give good fits to the radio data, but do not fully take account of what is known about galaxy populations and their evolution from multiwavelength data. In this work, we model the source counts using an infrared (IR) luminosity function as the base, exploring the relationship between radio and IR emission.

In Rowan-Robinson (2001, hereafter Paper 1), a source count model using a parametrized approach to galaxy evolution was presented. This method models the luminosity evolution, dominated by the star formation history (SFH), as a combined power law and exponential, the relative strengths of which are allowed to vary freely. This method is capable of reproducing most physically realistic forms of the SFH and allows a large set of possible histories to be investigated. The best-fitting parameter range is then determined by comparison with source count data. This parametrized approach gives good fits to the data in the submillimetre, far- and mid-infrared regions of the spectrum ($10 \umu$m $< \lambda < 1250 \umu$m), but underpredicts the optical and near-infrared counts. It was shown in King \& Rowan-Robinson (2003, hereafter Paper 2) that this issue could be resolved by introducing weak density evolution and a variation of optical depth with redshift. This model gives good fits for source count data across the spectrum, from UV to millimetre.

Extending this source count model to the radio region of the spectrum allows us to investigate the relationships between emission in the radio and other wavebands for different types of galaxies and the nature of radio sources. This paper presents this extension, which is done by extrapolating the spectral energy distributions (SEDs) to the radio using known infrared-radio correlations and introducing a minimal number of new assumptions. 

The structure of this paper is as follows. In Section \ref{sec:scmodel}, a brief explanation of the source count model presented in Papers 1 and 2 is given. In Section \ref{sec:normal}, the connection between infrared and radio for non-AGN galaxies is explored. Section \ref{sec:agns} looks at radio emission from quasars. The radio source counts are presented in Section \ref{sec:counts}. The implications of this model are discussed in Section \ref{sec:discussion}. Finally, the conclusions are presented in Section \ref{sec:conclusions}.


\section{Parametrized source count model}
\label{sec:scmodel}

Four basic ingredients are required to construct a source count model: a luminosity function, a cosmology, a spectral energy distribution (SED) and a form of evolution of the luminosity function and SED with redshift. The model used in this work is based on that presented in Paper 1 and expanded in Paper 2. The key features of this model are presented below; a more detailed description of the model can be found in Papers 1 and 2. A Lambda cosmology ($\Omega_{0} = 0.3$, $\Lambda = 0.7$) is used throughout, with $H_{0} = 100$ kms$^{-1}$Mpc$^{-1}$.


\subsection{Luminosity Function}
\label{subsec:lumfunc}

The luminosity function used is determined from the {\it IRAS} PSCz sample (Saunders et al. 2000). This data is then fitted with the empirical form of Saunders et al. (1990): 

\[
 \eta(L) = \frac{d\Phi}{d\lg L} =
 C_{*} \left( \frac{L}{L_{*}} \right)^{1-\alpha} \exp \left[ - \frac{1}{2\sigma^{2}} \lg^{2} \left( 1+\frac{L}{L_{*}} \right) \right], 
\]
where $C_{*}, \alpha$ and $\sigma$ are constants. For $\alpha$ and $C_{*}$, fixed values of $\alpha=1.09$ and $C_{*} = 0.027(H_{0}/100)^{3}$Mpc$^{-3}$ give the best fit to the luminosity function data. As the evolutionary parameters are changed, the other luminosity function parameters have to be adjusted to maintain consistency with the PSCz data. For the best-fitting Lambda model of Paper 2, they are $\sigma = 0.604$ and $\lg L_{*} = 8.688 L_{\sun}$. These parameters give the best fits to the Saunders et al. (2000) data. Physical descriptions of the parameters are given in Saunders et al. (1990).


\subsection{Parametrized form of Evolution}
\label{subsec:evolution}

In the model presented in Paper 1, it is assumed that changes in star formation rate (SFR) dominate the luminosity evolution, and this is parametrized to allow a wide range of model star formation histories to be explored. The form of the star formation rate, $\dot{\phi_{*}}(t)$ (in units of M$_{\odot}$ yr$^{-1}$ Mpc$^{-3}$), adopted is:

\begin{equation}
 \label{eqn:sfh}
 \frac{\dot{\phi_{*}}(t)}{\dot{\phi_{*}}(t_{0})} = \left[\exp\:Q\left(1-\frac{t}{t_{0}}\right)\right]\left(\frac{t}{t_{0}}\right)^{P},
\end{equation}
where $P$ and $Q$ are positive variable parameters.

Paper 1 assumes, for computational purposes, that $\dot{\phi_{*}}(t) = 0$ for $z > 10$. This is somewhat lower than recent {\it WMAP} polarization results, which give a reionization age closer to $z \sim 20$ (Kogut et al. 2003), but this is found to have a negligible effect on the predicted counts.

In a hierarchical galaxy-formation scenario, it is expected that a lower total density of galaxies will be seen at low redshifts as a result of merger events. Paper 2 explores this density evolution in the form:

\begin{equation}
 \label{eqn:density}
 \rho(z) = \rho(0)(1+z)^{n},
\end{equation}
along with a variation of optical depth with redshift. 

The best-fitting model of Paper 2 has, for a Lambda Universe, $P=3.4, Q=9.0, n=1.1$.


\subsection{Spectral Energy Distributions}
\label{subsec:seds}

Paper 1 explored a variety of assumptions about the SEDs. The method that was most consistent with available data was to derive the counts separately for each of four components -- M82-like starburst, `cirrus' (normal, non-active galaxy), Arp 220-like starburst and AGN -- then sum them.

For the starburst and cirrus components, the SEDs in the IR are based on the radiative transfer model of Efstathiou, Rowan-Robinson \& Siebenmorgen (2000). For the AGN component, the dust torus model of Rowan-Robinson (1995) is used. The SEDs, including the extension to the radio, are shown in Figs. \ref{fig:normsed} and \ref{fig:rlqsed}.

The proportions of the four components at 60-$\umu$m as a function of luminosity were chosen to give correct relations in colour-colour diagrams, as detailed in Paper 1. The relative proportion of the 60-$\umu$m emission due to AGNs is derived from the 12-$\umu$m luminosity function of Rush, Malkan \& Spinoglio (1993). We use the same functions as are listed in table 1 of Paper 1.


\section{Galaxy SEDs}
\label{sec:normal}

Observations of star-forming galaxies (starbursts and ULIRGs) show a very strong and well-established correlation between their FIR and radio fluxes, covering over four decades of IR flux intensities (e.g. de Jong et al. 1985; Helou, Soifer \& Rowan-Robinson 1985). This correlation has been also been observed in a wide variety of normal (non-AGN) galaxies, such as ellipticals and irregulars (e.g. Wunderlich, Wielebinski \& Klein 1987)

The radio and FIR fluxes are correlated in these normal galaxies because they arise from (different) physical processes that correlate directly with star formation (Helou \& Bicay 1993). There are two main sources of radio emission from normal galaxies (see review by Condon 1992): 

\begin{enumerate}
  \item  non-thermal synchrotron emission from relativistic electrons (cosmic rays) accelerated by supernovae remnants,
  \item  thermal free-free emission from H\textsc{ii} regions ionized by massive stars.
\end{enumerate}

The relativistic electrons produced by supernovae remnants have lifetimes on the order of $10^{8}$ years, while the free-free emission lasts $\sim 10^{7}$ yrs. As star formation occurs on a timescale of around $10^{7}$ years, the radio emission, like the FIR emission, is strongly linked to the star formation rate. For normal galaxies, the thermal free-free emission is typically one or two orders of magnitude less than that of the non-thermal synchrotron emission.

This correlation makes it possible to determine the counts of the non-AGN galaxies in the source count model by extending their SEDs to the radio.


\subsection{Cirrus SED} \label{subsec:cirrussed}

The radio SED for the cirrus component can be determined from the FIR using the known radio-FIR correlation. Helou et al. (1985) express this correlation as:

\begin{equation}
	\label{eqn:q}
	q = \lg \left[\frac{S_{FIR}/(3.75 \times 10^{12}\:\textrm{Hz})}{S_{\nu}(1.4\:\textrm{GHz})}\right],
\end{equation}
where $3.75 \times 10^{12}\:$Hz is the frequency at 80 $\umu$m and $S_{FIR}$ is an estimate of the flux between 42.5 and 122.5 $\umu$m:

\[
	S_{FIR} = 1.26 \times 10^{-14} \times [2.58\,f_{\nu}\:(60 \umu m) + f_{\nu}\:(100 \umu m)],
\]
where $f_{\nu}$ are the flux densities in Jy, and $S_{FIR}$ is in Wm$^{-2}$.
 
For the cirrus component, $q$ = 2.48 is used, from the value derived for spiral galaxies in Roy et al. (1998). Putting the 60 and 100 $\umu$m values in from the cirrus SED into eqn. \ref{eqn:q} gives a flux value at 1.4 GHz.

\begin{figure}
 \epsfysize=7.5cm
 \epsfxsize=8.0cm
 \epsffile{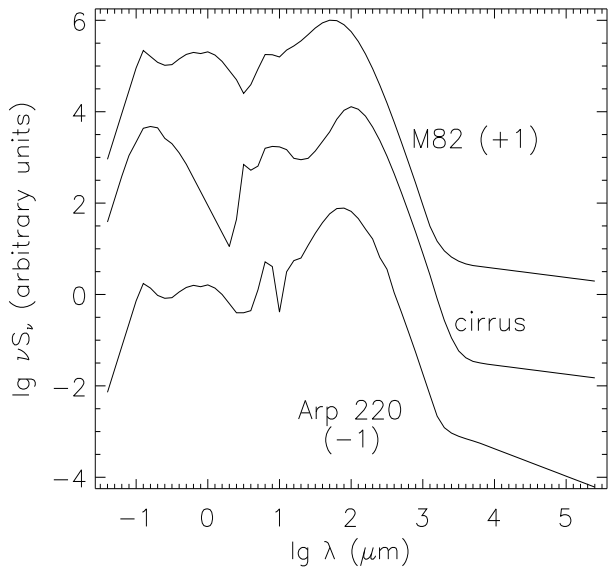}
 \caption{Spectral energy distribution for non-AGN galaxies. The top curve is the M82 starburst component (using $q=2.21$ and $\alpha = 0.8$), offset by +1, the middle is the cirrus (using $q=2.48$ and $\alpha = 0.8$) and the bottom curve is the Arp 220 starburst component (using $q=2.36$ and $\alpha = 0.4$), offset by -1.}
\label{fig:normsed}
\end{figure}

For normal galaxies, the radio continuum is expected to be dominated by synchrotron emission following the relation:

\[
	S_{\nu} \propto \nu^{-\alpha},
\]
where $\alpha$ is the spectral slope, having different values for different galaxy types. The spectral slopes of cirrus galaxies are found to be tightly correlated around a mean value of 0.8 (Gioia, Gregorini \& Klein 1982; Niklas, Klein \& Wielebinski 1997), with a standard deviation of 0.02. Using this, and the flux at 1.4 GHz derived from the FIR-radio correlation, the SED can be extended to the radio, using a spline fit to connect in the millimetre region. The resulting SED is shown in Fig. \ref{fig:normsed}.

\subsection{M82 starburst SED}

The starburst component can be extended to the radio using the same procedure as for the cirrus component (\S\ref{subsec:cirrussed}). The value of $q$ in eqn. \ref{eqn:q} is very similar to that for the cirrus component: Yun, Reddy \& Condon (2001) find $q=2.21$ for starburst galaxies. The mean spectral slope is the same as for cirrus galaxies, $\alpha = 0.8$ (Gioia et al. 1982; Niklas et al. 1997). These were used to extend the SED as shown in Fig. \ref{fig:normsed}.


\subsection{Arp 220 starburst SED}

The same procedure used for the cirrus and starburst galaxies can be used again for the Arp 220 type starbursts (ULIRGs). For these, Farrah et al. (2003) find the mean value of $q$ to be 2.36. However, the spectral slopes are more varied than for cirrus and starburst galaxies, ranging from 0.0 to 0.8 (Condon et al. 1991; Sopp \& Alexander 1991b), possibly because some ULIRGs contain AGNs as well as starbursts (e.g. Sanders et al. 1988; Genzel et al. 1998; Farrah et al. 2001). A mean value of 0.4 is adopted. The resulting SED is shown in Fig. \ref{fig:normsed}.


\section{Quasars}
\label{sec:agns}

It has long been known that only a small percentage of AGNs emit a significant fraction of their bolometric power in the radio region. The cause of this is still a matter of debate. Some studies have shown the distribution of radio luminosities to be bimodal (e.g. Miller, Peacock \& Mead 1990; Stocke et al. 1992; Hooper et al. 1995), prompting many to classify quasars as either `radio loud' or `radio quiet'. 

A number of suggestions have been put forward to explain the apparent division of quasars into radio loud quasars (RLQs) and radio quiet quasars (RQQs), such as differences in geometry (jets pointing towards the observer), absorption (a local, clumpy plasma), fundamentally different types of objects, different black hole properties, such as mass or spin (Lacy et al. 2001; Wilson \& Colbert 1995, respectively) and the presence or absence of lobes (see review by Antonucci 1993). The last two theories are generally considered to be the most likely to explain the full range of observations. One would expect that, if black hole properties were the cause, a continuum of radio loudness would be seen (e.g. Condon et al. 1981; Lacy et al. 2001), whereas the lobes hypothesis would produce two distinct populations. 

To include AGNs into the source count model in the radio region, it has been assumed that the presence or absence of lobes is responsible for the difference in radio emission, hence there are two populations (radio loud and radio quiet) with different SEDs.


\subsection{Radio quiet quasar SED}

The correlation between FIR and radio holds for all normal (non-AGN) galaxies, as both forms of emission are indirectly due to star formation (as discussed in \S\ref{sec:normal}). It is not immediately obvious, however, that there should be any link between FIR and radio for galaxies with AGN. However, Sopp \& Alexander (1991a) found that RQQs follow the same correlation as normal galaxies, although they show more scatter (Roy et al. 1998).

The most probable cause of this correlation is that the radio emission from RQQs is dominated by star formation activity in the host galaxy, not the AGN (Roy et al. 1998). It is most likely that the host is undergoing a starburst, causing both the FIR and radio emission. 

It has been shown that radio emission from quasars is split between the host galaxy (`disc') and a core, with the emission from RQQs being disc-dominated, and emission from RLQs being core-dominated (e.g. Sanders \& Mirabel 1985). This supports the hypothesis that radio emission from RQQs originates in the host galaxy, whereas RLQ emission is AGN-dominated.

In the model presented in Paper 1, the AGN component does not include the emission from the host galaxy, which is accounted for in the starburst component. Hence, the direct radio emission from RQQs is considered to be negligible.


\subsection{Radio loud quasar SED} \label{subsec:radrlq}

Unlike the RQQ case, radio emission from RLQs is core dominated, so there is no clear reason to expect a correlation between the FIR and the radio as the AGN SED peaks in the mid-IR. Empirical studies have, however, shown that a loose correlation exists (e.g. Sopp \& Alexander 1991a). This might be expected  if both the AGN and starburst share a common trigger event or fuelling mechanism (discussed in \S\ref{sec:discussion}).

If both the AGN and the star formation in the host galaxy are caused by merger or interaction events, then it is reasonable that the FIR and radio emissions should show some correlation in RLQs. This correlation, however, is much less tight and less well studied than for normal galaxies and RQQs, so four trial SEDs were created, based on two quasar catalogues: Elvis et al. (1994) and Poletta et al. (2000). From these catalogues, mean differences in flux between 60 $\umu$m and 1.4 GHz were derived. These were then used to determine 1.4 GHz flux values for each model from the 60 $\umu$m flux of the existing AGN SED. 

\begin{figure}
 \epsfysize=7.5cm
 \epsfxsize=8.0cm
 \epsffile{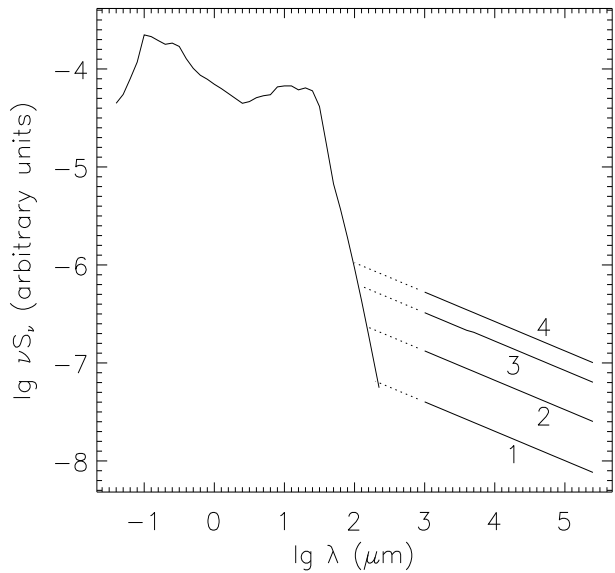}
 \caption{Four models of the spectral energy distribution of radio loud quasars, extended to the radio. The large variation in radio power is a reflection of the uncertainty of the FIR-radio correlation for RLQs. Note that the region shortwards of 1 mm (dotted lines) are not included in the model as little data is available and this part of the SED will have little impact on the radio counts.}
\label{fig:rlqsed}
\end{figure}

Quasars show a range of spectral slopes, $\alpha$, from 0 to 1, roughly dividing into two broad populations: flat spectrum ($\alpha \sim 0$) and steep spectrum ($\alpha \sim 1$). The mean slop is $\sim 0.7$, as steep spectrum quasars are more numerous. To simplify the model, it is assumed that all RLQs have this mean spectral slope. The same method as in \S\ref{subsec:cirrussed} is used to generate the SED, with $q = 2.48$ and $\alpha = 0.7$ (see Fig. \ref{fig:rlqsed}). The large range in variation of resulting radio flux reflects the uncertainty in the definition of RLQs and the scatter in the correlation between their FIR and radio emission.

Models were also tried where the spectral slope was varied between 0.5 and 1.0, but the effects on the resulting source counts in the data range examined were negligible. This indicates that the model would not be improved by splitting the RLQs further into flat and steep spectrum sources, or using a spread of values of spectral slope. This effect would be more noticeable at very long wavelengths and will be examined in future models.

\subsection{Radio loud quasar fraction}
\label{subsec:frrlq}

As RLQs are the dominant contributors to the bright counts, the treatment of the fraction of AGNs that are radio loud is very important to the model.

Early studies of the fraction of RLQs typically showed that around 10 to 20 per cent of quasars are radio loud (e.g. Kellermann et al. 1989). Models with a fixed fraction of AGNs being radio loud, however, do not come close to fitting the source counts at 1.4 GHz (see Fig. \ref{fig:radcount1}).

A number of studies have shown that the fraction of AGNs that are radio loud ($f_{RLQ}$) is not, in fact, fixed, but rather varies with redshift (more RLQs generally being seen at low redshift) and optical luminosity (e.g. Miller et al. 1990; Padovani 1993; Goldschmidt et al. 1999; Impey \& Petry 2001) - the two often being hard to separate. The cause of this variation of $f_{RLQ}$ is still poorly understood. Speculations on the cause include different rates of evolution and changes in opacity (Miller et al. 1990; La Franca et al. 1994).

\begin{figure}
 \epsfysize=7.5cm
 \epsfxsize=8.0cm
 \epsffile{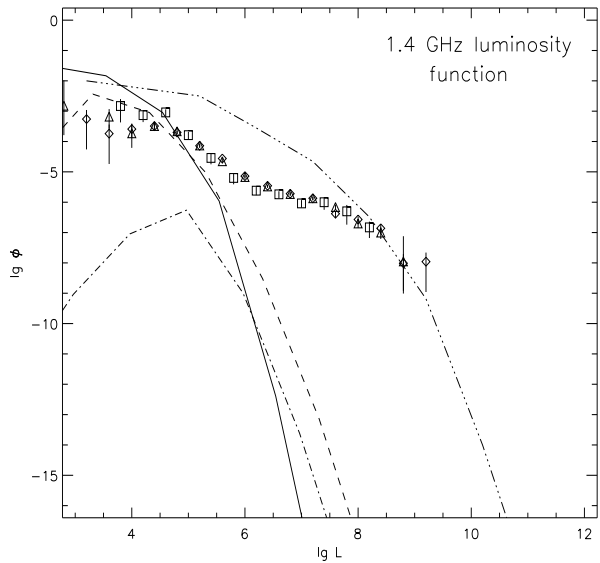}
 \caption{Local luminosity function at 1.4 GHz. Data are from Sadler et al. (2002) [diamonds], Machalski \& Godlowski (2000) [triangles] and Condon (1989) [squares]. The curves are the cirrus component [solid], M82 starburst [dashed], Arp 220 starbursts [dash-dot] and RLQ [dash-triple dot]. As can be seen, translating the FIR luminosity functions directly to the radio does not give the correct shape.}
\label{fig:radlf1}
\end{figure}

An investigation into the effects allowing $f_{RLQ}$ to vary with redshift showed that it could fit the radio counts reasonably well. However, there is a significant problem with this approach: it doesn't account for the known shape of the radio luminosity function (RLF). The RLF has a flatter shape than the IR luminosity function and simply translating the 60 $\umu$m luminosity function to radio using the SEDs thus far developed does not give the correct shape (see Fig. \ref{fig:radlf1}). Varying $f_{RLQ}$ with redshift will not affect the shape of the local RLF, so this mechanism does not fit all available data.

Allowing $f_{RLQ}$ to vary with luminosity could potentially solve the problem with the shape of the RLF by reducing the number of RLQs seen at fainter luminosities. It has been found that quasars with higher optical luminosities are more likely to be radio loud (Miller et al. 1990; Hooper et al. 1995; Goldschmidt et al. 1999; Impey \& Petry 2001). 

Data on the variation of $f_{RLQ}$ with optical luminosity (hereafter referred to as $f_{RLQ}(L_{B})$) is, like the variation with redshift, quite scarce and based on small samples. Hence, a number of models were constructed with $f_{RLQ}(L_{B})$ within the ranges presented by the available data and compared to the source count and luminosity function data to get a best-fitting model.

The best-fitting model using this approach was obtained using $f_{RLQ}(L_{B})$ as shown in Fig. \ref{fig:frlopt} and with a slight increase in the AGN luminosity function at luminosities $\lg (L/L_{\sun}) \sim 10 - 11$. This alteration in the luminosity function was sufficiently small to have no noticeable effect on the blue quasar counts or the 12 $\umu$m luminosity function. The resulting local luminosity function is shown in Fig. \ref{fig:radlf2}.


\section{Radio counts}
\label{sec:counts}

\begin{figure}
 \epsfysize=7.5cm
 \epsfxsize=8.0cm
 \epsffile{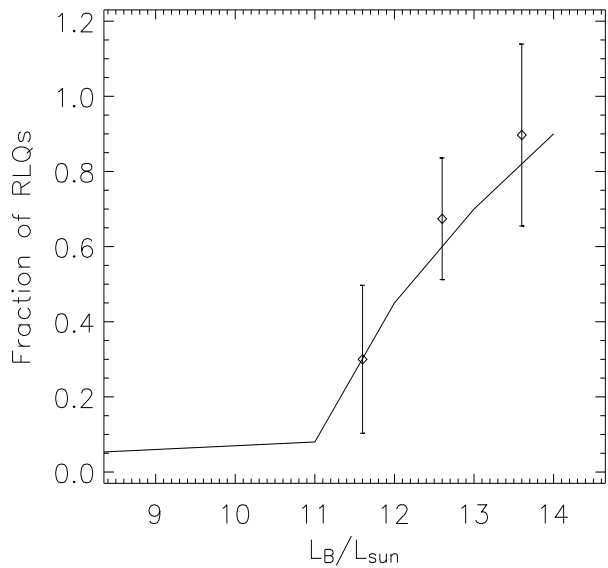}
 \caption{Best-fitting variation of fraction of radio loud quasars with optical luminosity. Data is from Impey \& Petry (2001).}
\label{fig:frlopt}
\end{figure}

\begin{figure}
 \epsfysize=7.5cm
 \epsfxsize=8.0cm
 \epsffile{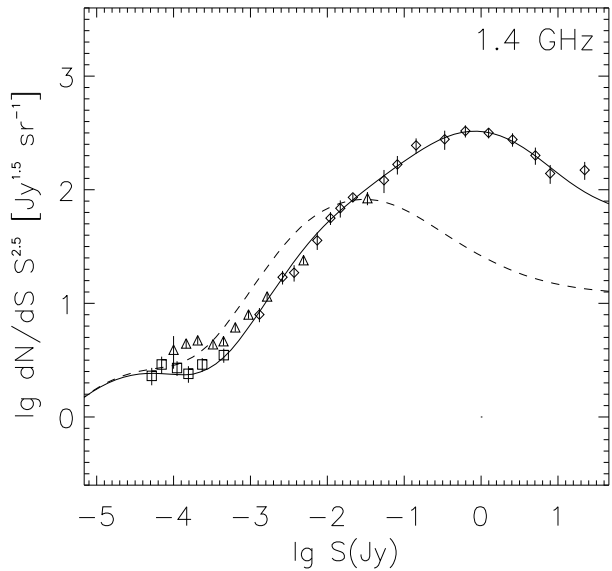}
 \caption{Euclidean-normalised differential source counts at 1.4 GHz. Solid line is the best-fitting Lambda model with $f_{RLQ}$ varying with optical luminosity and SED 4. This method gives a good fit to the counts, as well as the local radio luminosity function. For comparison, a model with a fixed $f_{RLQ}$ of 15 per cent and SED 3 is shown [dashed-dotted line]. Data are sub-samples of Richards (2000) [squares], Hopkins et al. (2003) [triangles] and Windhorst et al. (1993) [diamonds].}
\label{fig:radcount1}
\end{figure}

The best-fitting model uses the variation of $f_{RLQ}(L_{B})$ shown in Fig. \ref{fig:frlopt} with SED 4 and matches the data well at 1.4 GHz, as shown in Fig. \ref{fig:radcount1}. This figure also shows the counts for a model with a fixed  $f_{RLQ}$ of 15 per cent, with SED 3, for comparison. All the counts shown are differential, and normalised to a Euclidean Universe.

\begin{figure}
 \epsfysize=7.5cm
 \epsfxsize=8.0cm
 \epsffile{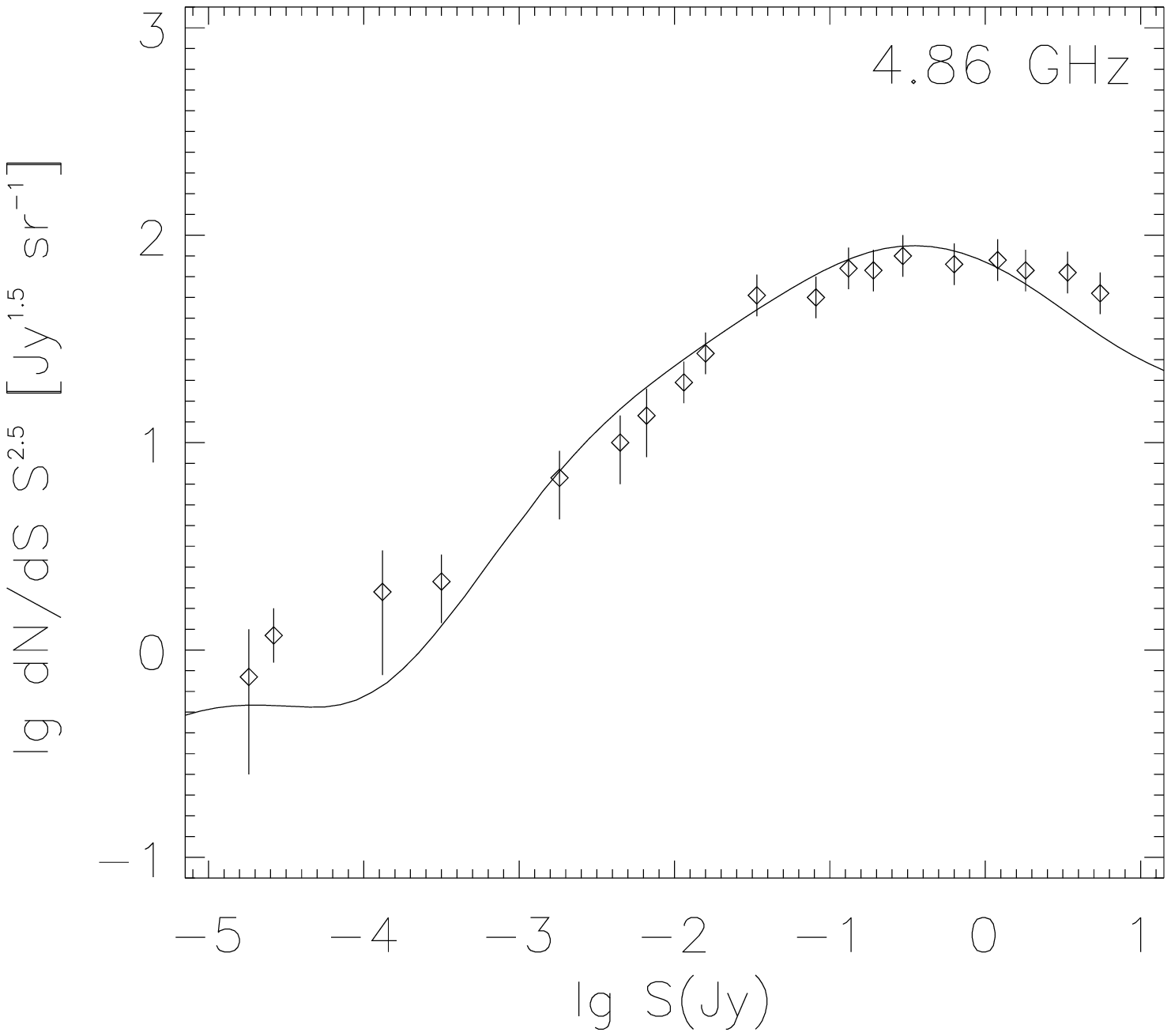}
 \caption{Best-fitting source count model at 4.86 GHz. The counts are differential, and normalised to a Euclidean Universe. Data from Windhorst et al. (1993) and references therein.}
\label{fig:radcount2}
\end{figure}

\begin{figure}
 \epsfysize=7.5cm
 \epsfxsize=8.0cm
 \epsffile{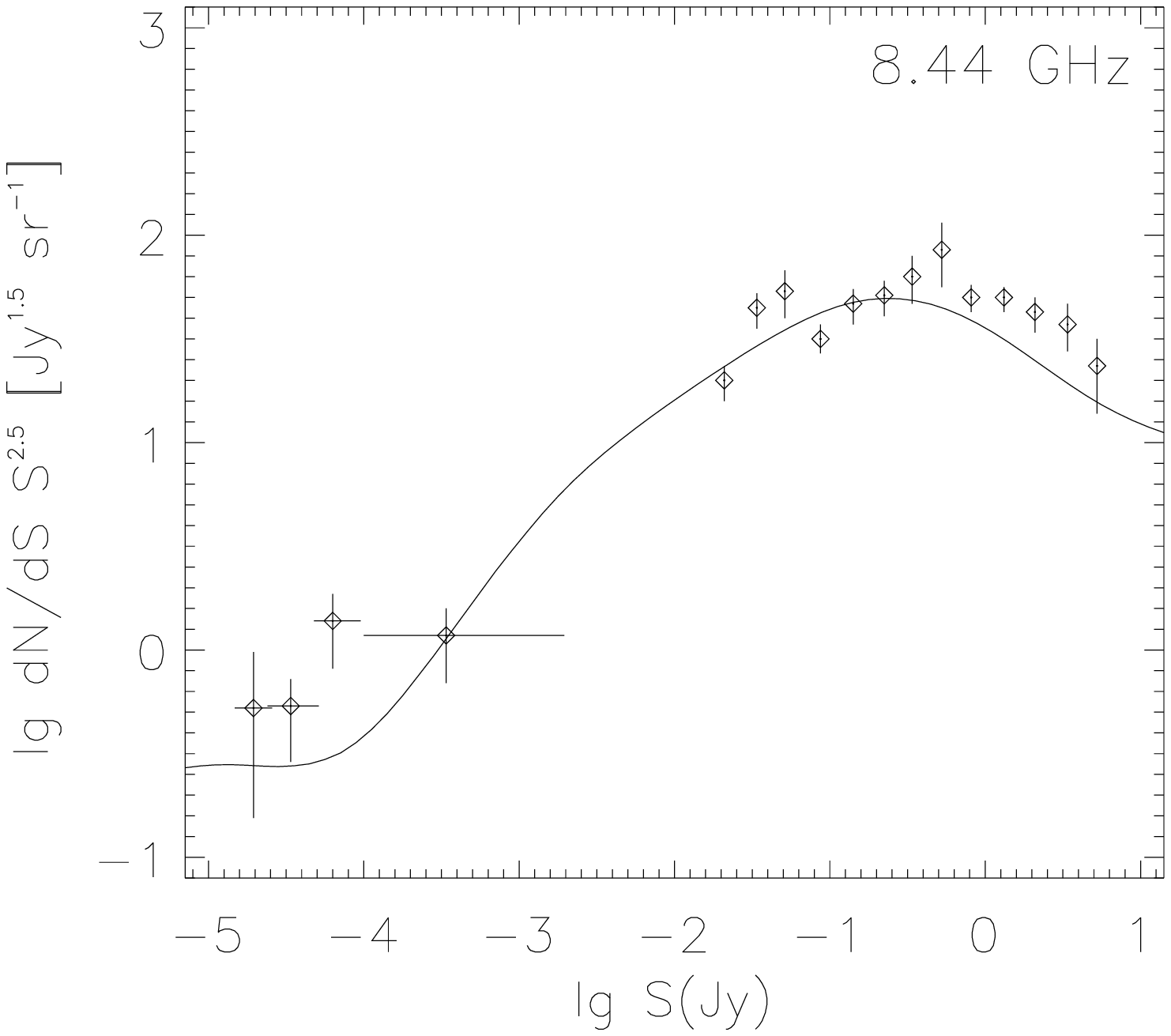}
 \caption{Best-fitting source count model at 8.44 GHz. The counts are differential, and normalised to a Euclidean Universe. Data from Windhorst et al. (1993) and references therein.}
\label{fig:radcount3}
\end{figure}

\begin{figure}
 \epsfysize=7.5cm
 \epsfxsize=8.0cm
 \epsffile{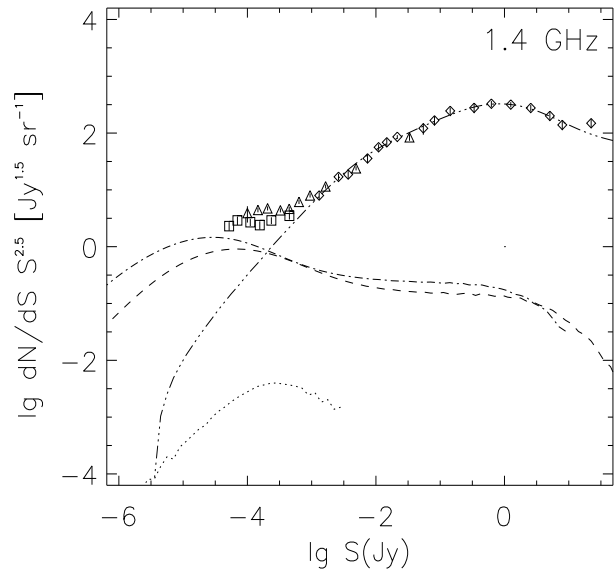}
 \caption{Best-fitting source count model at 1.4 GHz split into components. The counts are differential, and normalised to a Euclidean Universe. The components are cirrus galaxies [dotted-dashed curve], starbursts [dashed curve], Arp 220 starbursts[dotted curve] and RLQs [dash-triple dot curve]. Data as in Fig. \ref{fig:radcount1}.}
\label{fig:radfraccount}
\end{figure}

Good fits are also obtained to the counts at 4.86 and 8.44 GHz (Figs. \ref{fig:radcount2} and \ref{fig:radcount3}, respectively). Fig. \ref{fig:radfraccount} shows the relative contributions from the different galaxy types. Note that, in this model, it is predicted that the contribution of AGNs to sub-$\umu$Jy sources is negligible, which is not necessarily correct. It is possible that some faint source contain both starbursts and weak AGNs, which are not accounted for in this model. This could also explain why the predicted faint counts are at the low end of the available data, though this may alternatively be an indication that the selected values of $q$ for the cirrus and/or starburst components are slightly too high.

\begin{figure}
 \epsfysize=7.5cm
 \epsfxsize=8.0cm
 \epsffile{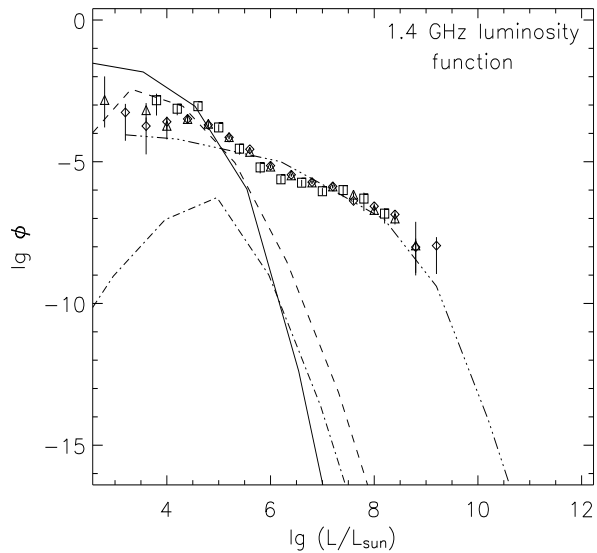}
 \caption{Local luminosity function at 1.4 GHz, using SED 4 for RLQs, with $f_{RLQ}$ varying with optical luminosity. Data as in Fig. \ref{fig:radlf1}. The curves are the cirrus component [solid], M82 starburst [dashed], Arp 220 starbursts[dash-dot] and RLQ [dash-triple dot]. This approach gives a reasonable fit to the radio luminosity function.}
\label{fig:radlf2}
\end{figure}

The fits obtained to the radio count and luminosity function data are generally good, though Fig. \ref{fig:radlf2} shows that the cirrus component is overpredicted at faint luminosities. It is possible that this may indicate a breakdown of the radio-IR correlation at low luminosities (Bell 2003).

\section{Discussion}
\label{sec:discussion}


\subsection{Relative numbers of RLQs and RQQs} 
\label{subsec:rlqrqq}

To explain the shape of the RLF, we found it necessary to classify AGNs as either radio loud or radio quiet, with a fraction that varies with optical luminosity, as is found observationally. This bimodality is essential for the model, so the possible reasons for it are worthy of discussion.

The best-fitting model (see Fig. \ref{fig:frlopt}) has a high fraction of RLQs at high optical luminosities ($f_{RLQ}=0.8$ at $\lg L_{B}=14$), dropping to a low fraction at lower optical luminosities ($f_{RLQ}=0.1$ at $\lg L_{B}=11$). Suggestions for the cause of this form have included RLQs having a slightly different evolution to RQQs and a difference in host galaxies. Other possibilities, which have been little discussed in the literature, are a variation in opacity and inverse Compton scattering of the microwave background.

\begin{enumerate}

\item
It is possible that AGNs might have a longer lifespan than starbursts, resulting in them persisting longer after the interaction or merger event. This difference might result in a different evolution, which Padovani (1993) showed could be responsible for the change in $f_{RLQ}$ with optical luminosity. The typical lifespans of both AGNs and starbursts, however, remain uncertain by several orders of magnitude.

A simple zeroth order estimate of timescale of AGNs, is the {\it e}-folding time, which is based on the black hole luminosity as a fraction of the Eddington luminosity. Very little, however, is known about this fraction, necessitating an empirical approach to the lifespan.

An upper limit on the lifespan of AGNs ($t_{AGN}$) can be obtained from the observed evolution of the quasar luminosity function (e.g. Osmer 1998), giving $t_{AGN} < 10^{9}$ yr, as the whole population rises and falls over this interval. Individual AGNs might exist on much shorter timescales than the population, however, and lower limits of $t_{AGN} > 10^{5}$ yr can be derived from indirect arguments, such as the requirement that quasars maintain their ionizing luminosity long enough to explain the proximity effect in the Ly$\alpha$ forest (e.g. Bajtlik, Duncan \& Ostriker 1988; Bechtold 1994). 

Estimates of the AGN lifetime, within these limits, are typically based on matching the present-day quasar luminosity functions to models of galaxy formation (Haehnelt, Natarajan \& Rees 1998; Salucci et al. 1999), or using measurements of high-redshift quasar clustering (Martini \& Weinberg 2001), yielding $t_{AGN} \sim 10^{6} - 10^{8}$ yr. 

The lifetime of starbursts, by comparison, is around $10^{6} - 2 \times 10^{7}$ years (e.g. Leitherer \& Heckman 1995; Thornley et al. 2000): a similar, though slightly lower, range to that found for AGNs. Given the uncertainty in these figures, it is not possible to definitely determine if there is any difference between the lifespans, though it cannot be ruled out.

\item
An obvious difference between RLQs and RQQs is the nature of their host galaxies, which could give their optical luminosity functions different shapes, resulting in the observed $f_{RLQ}(L_{B})$. Low redshift studies of AGN have long found that radio loud AGNs are more likely to be found in elliptical hosts, whereas radio quiet AGNs are found most often in spiral hosts. At higher redshifts, however, this distinction may break down (McLeod \& Rieke 1995; Taylor et al. 1996). Taylor et al. (1996) found that the fraction of RQQs in elliptical hosts increases with optical luminosity. This could produce a variation of $f_{RLQ}$ by affecting the optical luminosity function.

\item 
The variation in radio loud fraction could possibly be related to the Magorrian relation between black hole mass and bulge mass for galaxies (Magorrian et al. 1998). A relation between optical luminosity and bulge mass is to be expected, and there could also be a relation between radio loud fraction and black hole mass (depending on the cause of radio loudness), so the Magorrian relation could result in a variation of radio loud fraction with optical luminosity. 

\item
The existence of so-called `red' quasars could complicate the interpretation of the RQQ/RLQ split. Webster et al. (1995) reported the discovery of a population of quasars that do not appear in optically-selected samples, which they interpret as evidence of heavy obscuration in the optical region. If the Kellermann et al. (1989) definition of radio loudness, based on the ratio of radio to optical luminosity, is used, then some quasars would be mis-classified as radio loud, due to the optical extinction. This population of heavily-obscured quasars could also affect the perceived fraction of AGNs that are radio loud by suppressing the number of radio quiet quasars observed at optical wavelengths. It is not known how significant an effect this is, as the size of the population of red quasars is not certain. 

\item
The radio emission from quasars is predominantly synchrotron radiation from relativistic electrons. The relativistic electrons will also interact with photons from the cosmic microwave background (CMB) via an inverse Compton (IC) scattering process (e.g. Felten \& Morrison 1966; Rowan-Robinson 1970). The upscattering of these photons to optical or even X-ray energies will `use up' the energy in the electrons, diminishing the radio emission. This suppression of the radio emission could lead to sources being observed as radio quiet, though they would otherwise be seen as radio loud. The upscattering effect has been detected in radio sources and clusters (e.g. Fusco-Femiano et al. 1999). This effect would drop off as $(1+z)^{4}$, so could result in a variation in $f_{RLQ}$ with redshift, though the number of quasars that have sufficiently weak magnetic fields for the effect to be significant is not known. This effect is unlikely to be significant as radio sources above $\umu$Jy fluxes are predominantly at $z < 1$.

\end{enumerate}


\subsection{Influence of mergers and interactions} 
\label{subsec:radmerge}

As described in \S\ref{sec:agns}, the extension to radio relies on there being a correlation between the radio and FIR emission from RLQs. The goodness of fit indicates these assumptions are reasonable. If starbursts and AGNs were to share common trigger events and fuelling mechanisms, respectively, then this could explain the radio-FIR correlation. This theory is supported by a number of other analyses:

\begin{enumerate}

\item
If the star formation and black hole accretion processes are really concomitant, we would expect the luminosity functions of starbursting galaxies and AGNs to have similar shapes. Comparative studies of the bolometric luminosities of luminous FIR galaxies, bright optical quasars (Schmidt \& Green 1983) and lower luminosity Markarian Seyferts (Soifer et al. 1987) have shown that a broad similarity exists.

\item
For elliptical galaxies, the mass of the central black hole is proportional to the mass of the bulge (Magorrian et al. 1998). This relation also holds for quasars (Laor 1998; van der Marel 1999). Hierarchical models of galaxy evolution have shown that this relation occurs naturally if a common trigger for AGNs and starbursts is assumed (Cattaneo, Haehnelt \& Rees 1999; Kauffmann \& Haehnelt 2000).

\item
Quasars typically occur in areas of high galaxy density (e.g. Yee \& Green 1987), which could indicate that interactions and mergers are triggering gas inflows, which allow the formation of starbursts and provide fuel for the central black hole.

\end{enumerate}


\subsection{Breakdown of the IR-radio correlation?} 
\label{subsec:breakdown}

Recent work by Bell (2003) has cast doubt on whether the FIR-radio correlation holds for faint ($\sim 0.01L_{*}$) galaxies. It has been found that low-luminosity galaxies are optically thin, so the FIR should trace only a small fraction of the star formation. If radio were a perfect star formation indicator, a curvature should be detectable in the FIR-radio correlation. No such curvature is seen, which is interpreted as showing that radio is not a perfect star formation indicator, and hence the observed FIR-radio correlation is a coincidence for faint galaxies.


\section{Conclusions}
\label{sec:conclusions}

\begin{itemize}

\item 
The radio source count and luminosity function data can be fit well using a model based on a FIR luminosity function with 4 basic galaxy types, with the radio SEDs determined by the FIR-radio correlation and the AGN population being subdivided into radio loud and radio quiet populations.

\item
It has been shown that including a variation of the fraction of AGNs that are radio loud is necessary. A variation with redshift can provide a good fit to the source counts, but not to the local luminosity function, whereas a variation with optical luminosity will fit both. In the model where fraction varies with optical luminosity, no variation with redshift is required. It is not yet known why $f_{RLQ}$ might vary with optical luminosity.

\item 
This model supports the theory that AGNs and starbursts have a common trigger event, such as a merger or interaction, though they may have different lifetimes.

\item
The predicted faint counts are at the low end of the available data, which may indicate a missing population of weak AGNs, or an under-prediction of the radio loudness of the cirrus and/or starburst components.

\item
Future work is particularly needed on the variation of fraction of radio loud quasars with optical luminosity, which has been shown to be a significant effect. Other areas of investigation are: a comparison with the assumption of a continuum of radio loudness; the timescales of AGN and starbursts; the redshift distribution of radio sources; the opacity of quasars; the existence and nature of weak AGNs; and whether the FIR-radio correlation breaks down at high redshifts. Deeper radio observations are also required to investigate the r\^ole of the star-forming galaxies.

\end{itemize}


\section*{Acknowledgments}

We are grateful for comments by Jos\'e Afonso, Duncan Farrah, Matthew Fox, Chris Pearson and the referee, Rogier Windhorst.



\bsp
 
\label{lastpage}

\end{document}